\documentclass[12pt]{article}

\usepackage{pdfpages}
\usepackage[utf8]{inputenc}
\usepackage{authblk}
\usepackage{graphicx}
\usepackage{amssymb}
\usepackage{amsmath,amsthm}
\usepackage{url}
\usepackage{color}
\usepackage{natbib}
\usepackage[defaultfam,light,tabular,lining]{montserrat} 

\usepackage{setspace}
\usepackage{caption}
\usepackage{parskip}
\usepackage[utf8]{inputenc}

\usepackage[top=0.8in, bottom=1.6in, left=1in, right=1in]{geometry}
\title{A causal evaluation of Bogota's cable car illustrates the transformative potential of mobile phone data for policy analysis}

\author[1*]{\small Elena Lutz}
\author[2]{Sam Heroy}
\author[1]{David Kaufmann}
\author[2*]{Neave O’Clery}

\affil[1]{Department of Civil, Environmental and Geomatic Engineering, ETH Zurich}
\affil[2]{Centre for Advanced Spatial Analysis, University College London}
\affil[*]{Corr. Author: elelutz@ethz.ch, n.oclery@ucl.ac.uk}

\date{\small \today}

\begin{document}

\maketitle

\begin{abstract}

Transport infrastructure is vital to the functioning of cities. However, assessing the impact of transport policies on urban mobility and behaviour is often costly and time-consuming, particularly in low-data environments. We demonstrate how GPS location data derived from smartphones — available at high spatial granularity and in near real time - can be used to conduct causal impact evaluation, capturing broad mobility and interaction patterns beyond the scope of traditional sources such as surveys or administrative data. We illustrate this approach by assessing the impact of a 2018 cable car system connecting a peripheral low-income neighbourhood in Bogota to the bus rapid transit (BRT) system. Using a difference-in-differences event study design, we compare people living near the new cable car line to people living in similar areas near planned stations of a future line. We find that the cable car increased mobility by approximately 6.5 trips per person per month, with most trips within the local neighbourhood and to the city centre. However, we find limited evidence of increased encounters between the low income cable car residents and other socioeconomic groups, suggesting that while the cable car improved access to urban amenities and quality of life, its impact on everyday socioeconomic mixing was more modest. Our study highlights the potential of mobile phone data to capture previously hard-to-measure outcomes of transport policies, such as socioeconomic mixing.

\end{abstract}

\footnotesize{Acknowledgements: We accessed commercial mobile phone data from Quadrant via a programme that aids access to data for researchers. Elena Lutz gratefully acknowledges financial support from the Early Career Grant of the 'Leading House for the Latin American Region'. Emily Robitscheck provided excellent research assistance. The funders had no role in study design, data collection, and analysis, decision to publish, or preparation of the manuscript.}

\clearpage

\section*{Introduction}

Policy impact evaluation is vital for effective policy design and informed investment decisions, offering insights into both short- and long-term effects of programmes and interventions. Such evaluations are key for planning and the design of future interventions for a wide range of stakeholders including governments, international organizations, investors, and  donors. An example of an ambitious infrastructure policy, this paper focuses on a transport intervention (in the form of a new cable car in Bogota) that aimed to reduce commute times and open up economic opportunities via easier accessibility to jobs. However, evaluation of such policies is costly and relatively rare due to a combination of factors including cost, possible lack of data or lack of knowhow \citep{jacob2015institutionalization, maxnathan}. In particular, quantitative impact evaluations typically use survey data or administrative data, e.g., census data, to capture mobility patterns of individuals. In the case of travel surveys, the data sample is usually very limited in terms of population coverage, is infrequently collected, and records participants’ recollections rather than realised mobility which may suffer from imperfect memory errors and social desirability bias. Here we propose to harness GPS location data, collected via opt-in smartphone apps, for policy impact evaluation. This type of data is commercially available at massive scale, can be cost-effective relative to surveys, and captures realised high spatial resolution mobility signatures in real time. 

While there is a growing literature using this data to describe urban mobility and segregation patterns \citep{gonzalez2008understanding, athey2021estimating, barbosa2021uncovering, moro2021mobility, xu2025moro}, there are very few existing studies aiming to combine causal inference techniques with mobile phone GPS data. Some recent examples include the use of historic zoning plans to quantify the impact of local amenity access on the share of city trips under 15 mins \citep{abbiasov202415}, the use of historic mortgage zoning maps to identify areas of persistent experienced segregation \citep{aaronson2025lasting}, and the use of randomness in trips to government car registration offices to assess the impact of fast food availability on consumption \citep{garcia2024food}. These studies take a cross-sectional approach but for evaluation it is generally desirable to observe effects before and after the policy was implemented. Here we propose to combine smartphone GPS data with panel (temporal) causal inference methods, such as difference-in-difference (DiD) regression - one of the most widely used methods for policy impact evaluation \citep[e.g.]{roth2023difference}. Intuitively, DiD compares a “treatment group”, i.e., individuals that are affected by the policy, to a “control group” consisting of similar individuals not affected by the policy. This approach is often suited to policies with a spatial dimension, such as transport policies. Combining GPS mobile phone data and DiD causal research design represents a potential step-change in our ability to assess infrastructure investments, potentially in close to real-time. 

We illustrate this approach by investigating the impact of the 2018 opening of a new cable car, part of a planned wider system of cable cars called the ‘TransMiCable’, in Bogota, Colombia. As illustrated in Figure 1, the new cable car connects a poor neighbourhood, located on a steep hill in the south of Bogota, to the city’s Bus Rapid Transit (BRT) network, reducing travel time to the nearest BRT station by about 80\% at no extra cost. We compare people living close to the first line of the cable car, which opened at the end of 2018, with residents close to a second planned line expected to open in 2026 \citep[see e.g.]{tsivanidis2023evaluating}. We hypothesise that the new cable car would impact the number and type of trips taken by local residents, and facilitate encounters with higher-income groups during their daily activities. 

Tackling rugged terrain and high levels of informality, Latin American cities have a long history of innovative and experimental transport solutions, including cable cars \citep{hidalgo2013implementation}. A large literature supports the idea that transport investments increase accessibility, shorten commute times, and improve socioeconomic outcomes, such as more leisure time and economic growth \citep{mackie2001value, donaldson2018railroads, asher2020rural, banerjee2023learning, tsivanidis2023evaluating, guzman2025lifeline}. A number of studies study the impacts of cable cars, including \cite{matsuyuki2020impact} who find that while Medellin’s cable car is not often used by the very poorest residents, local women did benefit from a shorter and safer commute. Focusing on the Bogota cable car, a survey found that residents perceived benefits from travel time savings and increased leisure time \citep{guzman2023expectations}, as well as increased pride in community and perceived security \citep{rubio2023impacts}.

Our key hypothesis is that residents who live close to the cable car will not only travel more frequently, but also visit different parts of the city and encounter new people. These encounters are crucial as they strengthen urban social networks \citep{chetty2022social}, drive support for redistributional tax policies \citep{sands2017exposure}, and decrease crime \citep{sampson2020crime}. We are interested in changes in ‘experienced segregation’, or the degree to which an individual meets other income groups during daily life, e.g. at parks or restaurants. An emerging literature on experienced segregation suggests that people tend to more frequently encounter others of the same race \citep{athey2021estimating} and class \citep{cook2024urban, heine2025role, moro2021mobility}. This segregation tends to be more intense in larger cities \citep{nilforoshan2023human} but lower in cities that are denser and have better public transit \citep{athey2021estimating}. Here we investigate the extent to which shortening the travel time between high-income and low-income neighbourhoods via public transport increases encounters across income classes (or equivalently decreases experienced segregation). 

To capture residents’ mobility patterns, we use location data (‘pings’) collected by opt-in smartphone apps (e.g., ride-hailing services or weather apps) to infer information at an individual level such as home location, trips outside the home and to specific amenities \citep{moro2021mobility}. We find that the opening of the cable car resulted in an average increase of approximately 6.5 trips (5\%) per person per month compared to the control group during the immediate 6 months after opening in December 2018, mostly within the local southeast region around the cable car and to the city centre. Local residents visited more amenities, including new amenities, and weakly increased their visits to amenities frequented by both diverse income classes and higher income classes. Taken together, this suggests that the cable car, consistent with survey evidence \citep{guzman2025lifeline}, benefitted residents in terms of increased local trips and activities, but did not lead to significantly greater exploration of the city or encounters with diverse income groups. 

\begin{figure}[t!]
    \centering
  \includegraphics[width=1\textwidth]{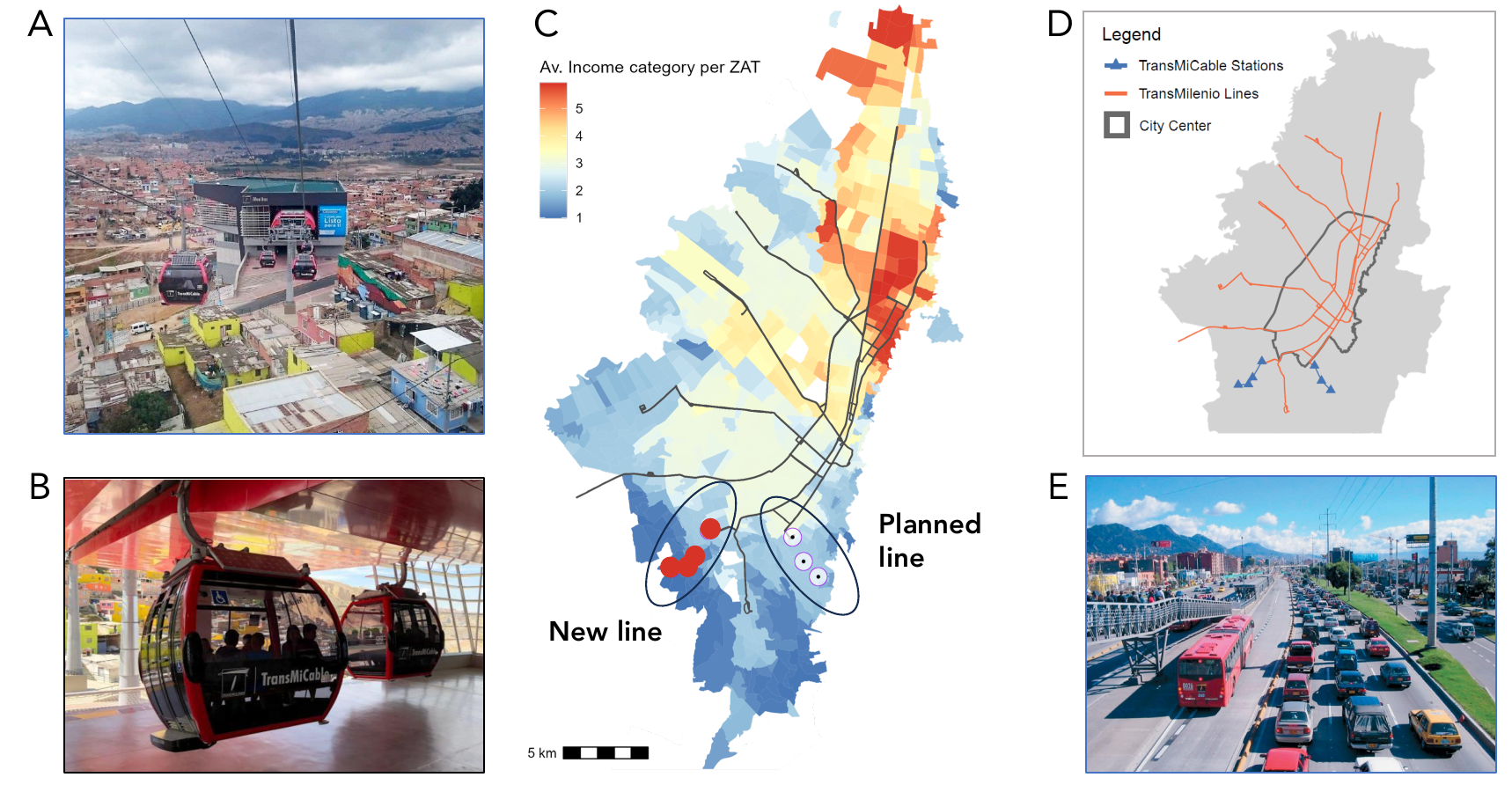}
  \caption{[A-B] The new cable car connects a poor area of Cuidad Bolivar, Bogota, to the Bus Rapid Transit (BRT) system. [C] There are two planned lines, one of which opened in 2018 and the other which is planned to open in 2026. [D-E] These new lines connect to the BRT which is a radial system of buses with their own dedicated bus lane that runs throughout the wider city.}
  \label{fig1}
\end{figure}

\section*{Results}

\textbf{GPS location data.} We deploy a difference-in-difference (DiD) study to causally infer the impact of the opening of the cable car on the mobility of residents living around its four new stations (the treatment group as illustrated graphically in Figure 2A). Specifically, we compare the mobility of residents living within 500m of the cable car stations to that of residents living near planned future cable car stations (the control group). We focus on the period 6 months before and after the opening of the cable car in December 2018 (so July 2018 to June 2019) during which our dataset contains 662m ‘pings’ (location time-stamps) from 902k mobile phones in Bogota. Following a commonly used algorithm \citep{hariharan2004toyama}, we detect 746k home locations (corresponding to ca. 10\% of the population), including approx 3,300/4,200 homes in the treatment and control area, respectively. The correlation at the neighbourhood (ZAT level) between the number of individuals according to official census data and the number of homes detected in the mobile phone data is 0.79 (see SI). For individuals in the treatment and control areas, we detect trips to locations outside the home (referred to as ‘stays’ in the technical literature) within the municipality of Bogota, as illustrated in Figure 2C, finding on average 100 trips per month per person. In the DiD analysis below, we group users into 261 hexagons (100m side) which lie within 500m (approx a 10-minute walk) of the treatment and control cable car stations. 

\begin{figure}[t!]
    \centering
  \includegraphics[width=1\textwidth]{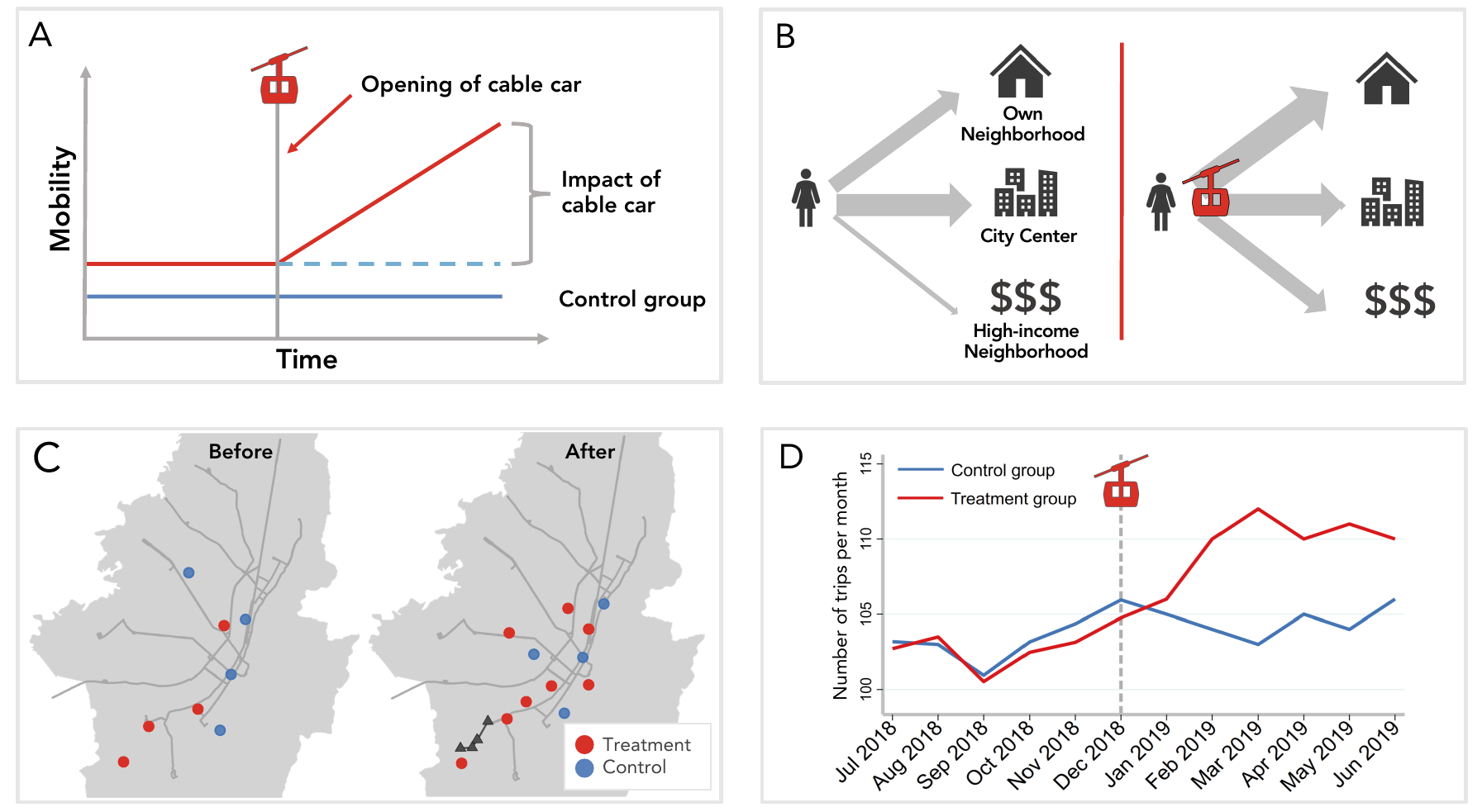}
  \caption{[A] Graphical illustration of our difference-in-difference (DID) event study. We wish to investigate the extent to which the cable car opening results in a divergence in mobility patterns of the treatment and control group after the cable car opens. [B] We hypothesise that the cable car might induce a change in visitation patterns, i.e., residents visit different parts of the city such as richer or more commercially-active areas. [C] We capture trips (and destinations) using mobile phone GPS 'pings' for residents of both the treatment and control group before and after the cable car opening. [D] The parallel trends assumption, we find that the number of trips taken by those in the treatment and control group diverge after the opening of the cable car.}
  \label{fig2}
\end{figure}

\textbf{The treatment and control group have similar mobility patterns pre-treatment.} A necessary condition for DiD event studies is that the time trend of the outcome variable (in our case the number of trips) is similar across both the treatment and control group before the intervention (the so-called ‘parallel trends’ assumption). In Figure 2D, we illustrate parallel pre-trends for the number of trips per person per month, with divergence in trends post cable car. Second, we also harness a 2019 mobility survey carried out by the City of Bogota (Encuesta de Movilidad) to compare the two areas across a range of socio-economic variables. Figure 3A-D shows that the treatment and control groups have similar occupational profiles, age distributions, income distributions and schooling levels. The treatment group skews slightly younger with more residents in the 0-9 age group and primary school categories. Figure 3E finds comparable rates of both phone ownership and mobile data plans, as well as a similar gender balance.

\begin{figure}[t!]
    \centering
  \includegraphics[width=1\textwidth]{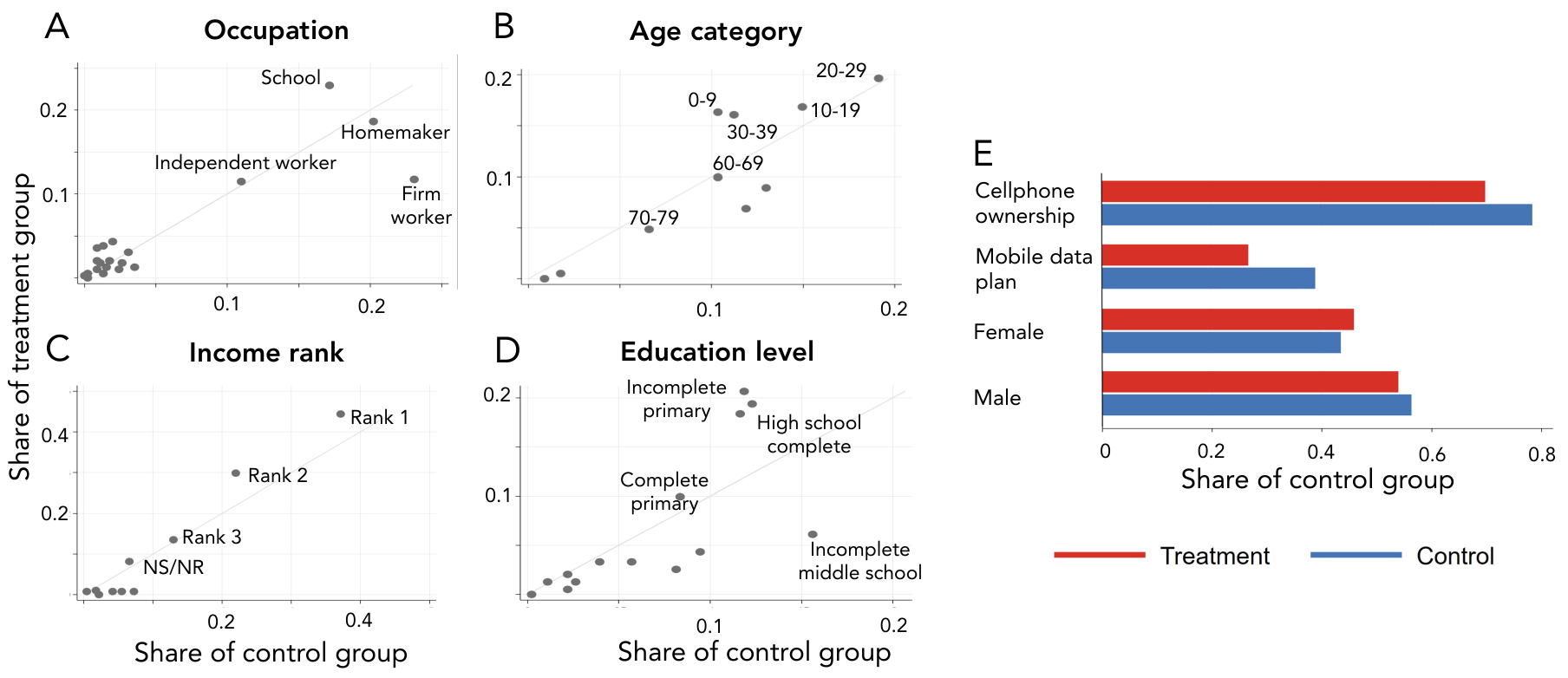}
  \caption{[A-D] The treatment and control groups have similar occupational profiles, age distributions, income distributions and schooling levels. [E] They also have similar rates of both phone ownership, mobile data plans, and gender ratio.}
  \label{fig3}
\end{figure}

\textbf{We detect an impact of the cable car on mobility patterns.} In order to evaluate the impact of the cable car on the number of trips a person makes in a month, we estimate the following model: 
$$
y_{i,t}=\sum_{t=jul 18:jun 19, \atop t\neq nov 18}\beta_t \text{ x }Cable_{i,t}+ \zeta_i + \zeta_t+\epsilon_{i,t}
$$
in which $y_{i,t}$ is the average number of trips outside the home per person per month in hexagon $i$ and month $t$, $Cable_{i,t}$ is a dummy equal to 1 if hexagon $i$ has a cable car in month $t$ and zero otherwise. In this case, $\beta_t$ can be interpreted as the estimated average difference in monthly trips between the treatment and the control group in month $t$ when controlling for hexagon and month fixed effects ($\zeta_i$ and $\zeta_t$). If there is a detectable causal impact, the value of $\beta_t$ will become significantly different from zero after the intervention. Figure 4A shows graphically the results of the full model. While we observe parallel trends before the opening of the cable car, we detect a sustained increase in the number of trips outside the home after the opening, with the treatment group making an estimated 6.5 (or equivalently 5 percent) more trips per person per month.

\textbf{Cable car residents travel locally and to the city centre.} Next, we break trips down by destination in the city. To infer socio-economic status of city neighbourhoods, we use an official classification of city blocks into so-called ‘stratum’. Based on visual characteristics, each block has been assigned a stratum level, with 1 being the poorest and 6 the richest. We group these into low (stratum 1-2), middle (stratum 3-4) and high (stratum 5-6) income areas. We find a robust statistically significant increase in trips to both low and middle income areas, but not high income areas. On average it seems that roughly half of new trips are to low income areas, and the other half to middle income areas. This initial finding suggests that while low income residents do, after the opening of the cable car, increasingly encounter middle income individuals throughout their daily activities, they have not increased travel to high income areas. 

To probe this further, we investigate the destination of trips, focusing on trips to different regions of Bogota separately, using the regions of the POT of the City of Bogota. We observe that, as shown in Figure 4E, the coefficient is only significant for two regions, the Center and the Southeast (where both cable car lines are). That these two regions see an increase in trips makes sense as the cable car links to the BRT which goes into the city centre. Local residents highlighted additional time to do leisure activities since the cable car as a key impact in a survey \citep{guzman2025lifeline}, which might explain the increase in local trips. We don’t find a statistically significant effect for trips to the other regions nor for the average distance travelled by cable car residents (see SI for both). The fact that residents don’t seem to increase visits to (or receive visits from) the wealthy North/Northwest, or other poor communities in the West and Southwest, suggests that the cable car has not increased exposure of low income cable car residents to higher income groups. We investigate this in more detail next.
\begin{figure}[t!]
    \centering
  \includegraphics[width=1\textwidth]{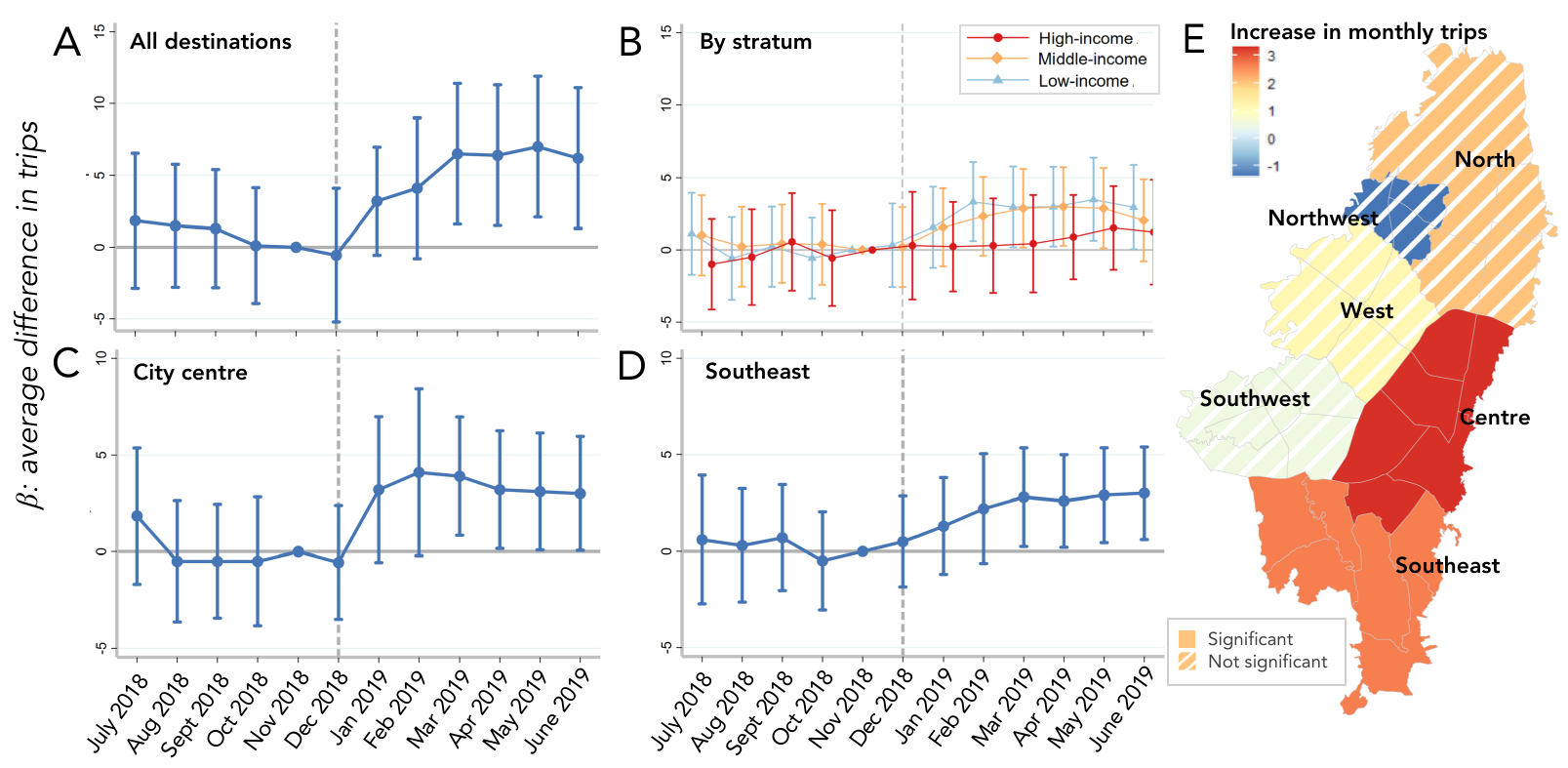}
  \caption{[A] We detect a sustained increase in $\beta_t$, the average difference in monthly trips between the treatment and the control group, after the opening of the cable car, with an average value of 6.5 stays per person per month. [B] This is true for trips to low (stratum 1-2) and middle (stratum 3-4) income areas, but not high income (stratum 5-6) areas. [C-E] We find a statistically significant increase in the number of trips to just two regions, the Center (CBD) and the Southeast (i.e., local trips). }
  \label{fig4}
\end{figure}

\textbf{Cable car residents benefit from more visits to amenities and exposure to diverse socio-economic groups.} First, we investigate whether the new cable car prompts an increase in visits to points of interest (POIs) such as shops, parks or schools (which might occur, for example, due to faster travel times or more leisure time). We collect POI locations from the Google Places API and assign a trip to a POI if it's within 50m (following \cite{moro2021mobility}, see SI for details). As shown in Figure 5A, we find a statistically significant increase in the number of POI visits, with an increase of approximately three visits per person per month. There is also a weak increase in the number of unique POIs visited, meaning that individuals in the treatment group tend to visit more different POIs. However, this effect is not statistically significant on any common significance level. 

Building on an extensive literature that studies ‘experienced segregation’ \citep{athey2021estimating, cook2024urban, moro2021mobility}, we are interested in measuring the extent to which cable car residents encounter others of different socio-economic status during their visits to POIs. To do this, for each individual POI, we quantify the diversity of the stratum of POI visitors, as illustrated in Figure 5C. Specifically, we compute the entropy of visits across stratum (income) groups (see Methods) - a higher entropy suggests that a POI is visited by people from a wider range of strata. Figure 5G suggests that high entropy POIs are located in the city centre, and also along a large transit corridor towards the airport in the north west of the city, which is home to many office blocks and government offices. As an alternative measure of exposure, we also compute the share of high income visitors to each POI, as illustrated in Figure 5D. Then, for each cable car resident (and those in the control group), we compute the average entropy and income share for the set of POIs they visit (before and after the cable car opening as above). Figures 5E-F show that while residents do appear to experience a modest increase in exposure to more diverse/high-income individuals at the POIs they visit, the effects are not robustly statistically significant at a 5\% level. This result is consistent with our analysis above that suggests that cable car residents benefit from more trips to the city centre, alongside local trips. 

\begin{figure}[t!]
    \centering
  \includegraphics[width=1\textwidth]{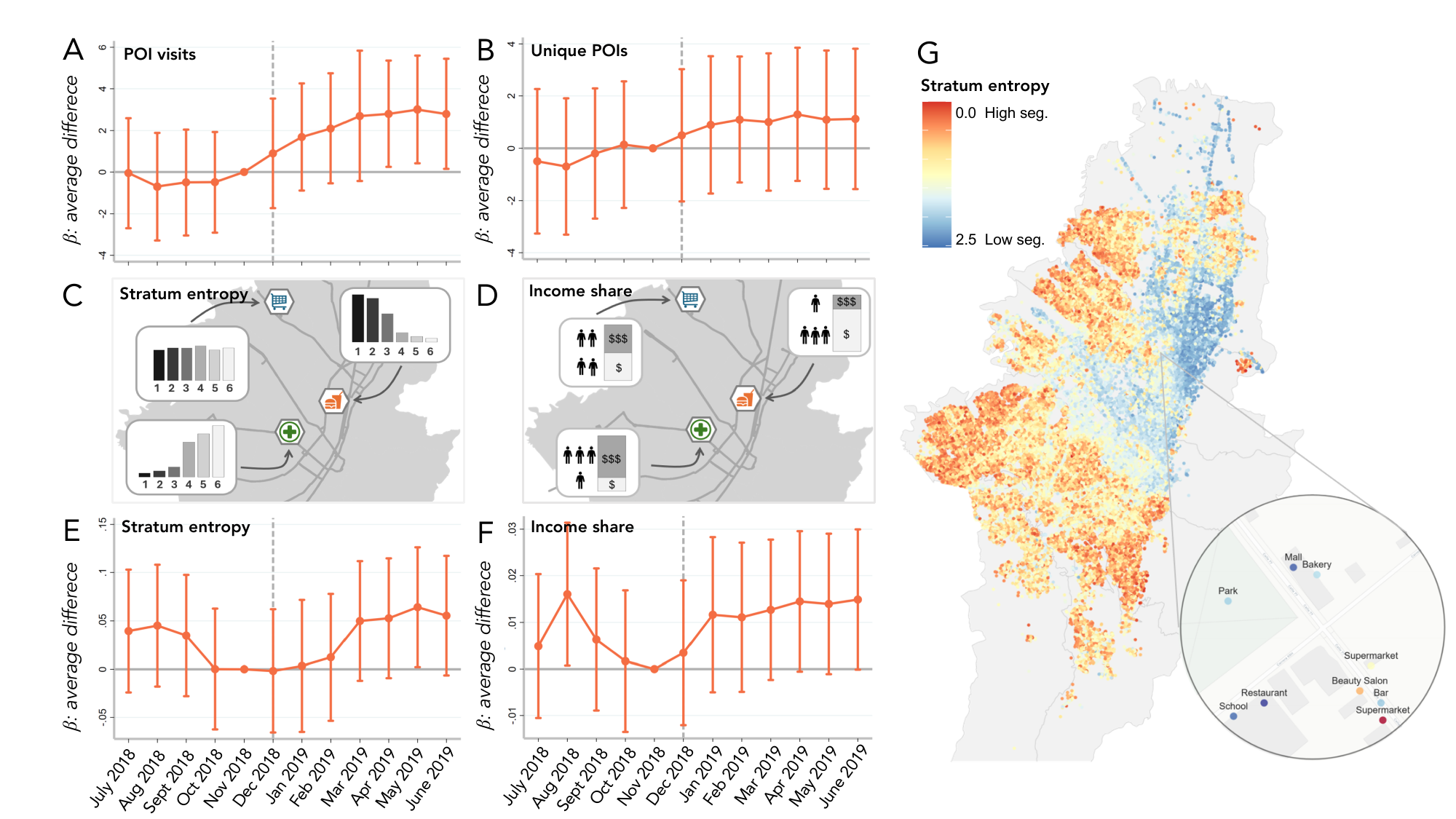}
  \caption{[A] We find an overall increase in trips to POIs for the treatment group relative to the control group. [B] There is a modest but not statistically significant increase in the diversity of POIs visited (measured as the number of distinct POIs). [C] To investigate the exposure of cable car residents to diverse income groups, we compute the stratum entropy (diversity of visitors across stratum groups) to each POI. [D] We also compute the share of high income visitors to each POI. [E] We observe that POIs with low entropy are located in the poorer south and west of the city, while POIs with a more even share of visitors across stratum groups (high entropy) are concentrated in the city centre and to the north and west of the centre. [F-G] We find a weak but non-significant increase in both visits to high entropy and high income POIs (relative to the control group) after the cable car opening.}
  \label{fig5}
\end{figure}

\section*{Discussion}

This paper illustrates the potential of using smartphone GPS data and methods of causal inference to evaluate the impact of opening a cable car in a low-income neighbourhood. We find that the cable car increased the mobility of individuals living close to its stations: they make significantly more trips outside the home in general and to POIs. Prior to the cable car, residents of the treatment area had limited spatial access to amenities and services, such as parks, hospitals, or restaurants, and areas of Bogota with a high POI density were difficult to reach. For example, reaching the city centre took ca. 2 to 3 hours resulting in very long commutes for many residents. Our study finds that the cable car enabled residents to make more local trips in their own neighbourhood, potentially strengthening local social networks, as well as visits to different places. Likely, both the travel time saved on commuting as well as the easier spatial access per se resulted in the observed increased mobility. However, these results only cover changes in mobility patterns in the first six months post-opening, missing potential structural changes over longer time periods. 

As Bogota is very segregated by income, and because the treatment area is peripheral and insecure, individuals must travel to other areas of the city to encounter other income groups. We find an increase in the number of trips made to middle-income areas in the city centre of Bogota, but not to more distant high-income areas in the north of the city. This may be partly because relative travel time reduction is greater for the city centre than the wealthy north. Consistent with the fact that Bogota’s city centre is visited by a wider range of socioeconomic groups, our results suggest a small increase in the economic diversity (Shannon entropy) and the share of high-income individuals that the treatment group encounters. This suggests that the opening of the cable car also opened up the potential for encounters with other income groups. Despite the fact that the effect size we observe is modest and not statistically significant at the 5\%  significance level, if sustained over time, this increased spatial exposure to other socioeconomic groups could help low-income individuals in the treatment group diversify their social networks \citep{chetty2022social}. However, it’s well established that POIs visited by an individual are also strongly determined by their financial resources and not only by spatial proximity \citep{boar2023consumption, davis2019segregated}, and so lowering spatial barriers is unlikely to be in itself sufficient for large changes in experienced segregation. Overall, our results suggest that policies which improve public transit access for very low-income individuals can increase their mobility and visits to urban amenities, but may struggle to significantly decrease experienced segregation, at least in cities with high residential income segregation, such as Bogota. 

A broader aim of this paper was to illustrate how widely available GPS location data from smartphones can be used to evaluate urban policy interventions. We focus on the example of a government opening a new public transit line, but the methodology can, in principle, be applied to a wide range of other urban policies and urban planning interventions, such as traffic slow zones or the opening of new parks. As smartphone GPS data is becoming more and more widely available, and methods for working with it more and more standardized \citep{wang2018applying}, this data holds great potential for evaluating the effects of urban policies. Even given the well-known drawbacks of this data, such as concerns about representativeness or privacy \citep{dypvik2021mobile}, big data from smartphones may be the only source of granular mobility data available in a majority of global cities. In addition, it is often available in real time, opening up the potential for monitoring impacts of policies as soon as they take effect. Similar to studies that show that big data from smartphones can be used to improve aid targeting \citep{aiken2022machine} or to measure wealth \citep{blumenstock2015predicting}, we illustrate that this data can be used to evaluate urban policies.

Yet, there are non-trivial limitations to this approach. First, to combine smartphone data with DiD designs, a suitable control group needs to be identified, such as another similar neighbourhood that is not affected by the policy, which is not always possible. Second, anonymized smartphone data, which usually does not contain any socio-demographic information, such as age, income, or gender, limits the outcomes researchers can study. For example, in our case, the data can be used to study mobility behaviour, but not whether there is a change in, e.g. income of treated individuals. We also cannot study gender differences in the usage of the cable car. The third limitation is that depending on the specific smartphone dataset, there can be limitations on the extent to which it's possible to follow individuals over long periods of time. For example, in our sample, there are many cellphone identifiers which we do not observe for the entire year but only for a few months. This indicates that while smartphone data can be used to conduct temporal studies, tracing single individuals over long periods of time - as it is possible e.g. using administrative registry data – may not be possible. Likely, the longer the time horizon, the more complex it is to trace individual cell phones. Thus, using smartphone GPS data combined with causal inference methods would ideally be complementary to using other traditional data sources rather than fully replacing them.

\section*{Methods} 

\subsection*{Data}

\textbf{Mobile phone data.} Our primary source of data, anonymous, privacy-compliant GPS mobile phone data, was provided through by Quadrant, a data analytics company. This project has full ethical approval from the ETH Zurich Ethics Commission (Project 25 ETHICS-029).   

The data spans one year from July 2018 to June 2019, covering six months before and six months after the cable car's opening. Each observation is a ‘ping’ characterised by a position expressed in latitude and longitude and a timestamp. The data set contains an anonymized cell phone identifier allowing us to assign each observation to a cell phone. Pings are of comparatively high spatial accuracy, with more granular accuracy (mean of 15.2 metres) than CDR data (nearest cell phone tower). The dataset is large covering ca. 4 billion observations. However, we exclude data from phones that are observed only infrequently (defined as phones with less than 50 pings per month or pings on less than 10 days in 2018 or 2019). 

\textbf{Mobility survey data.} We use a mobility survey called Encuesta de Movilidad 2019 from the DANE. The survey collects information by going door to door, and is representative at the ZAT (large neighbourhood) geographic level. Respondents are asked about their mobility within the last 24 hours, and information on socioeconomic characteristics such as age, education and income is collected. We use this data for three purposes: (1) it provides more detailed socio-demographic information on our study area, (2) and specifically on the demographics of smartphone owners in our areas of interest, and (3) we compare our mobility results to those captured by the travel survey. 

\textbf{Points of interest (POI).} We downloaded data on 40 different place categories from the Google Places API (see list in SI). The data includes many different consumption amenities, such as shops or restaurants, as well as parks, educational institutions, health-related places, places of worship or museums. In total, we obtained the location and type of 108,881 operational POIs. 

\subsection*{Measuring mobility and segregation}

\textbf{Detecting home locations and trips.} Next, we detect individuals' home locations, defined as the place where an individual stays most often at night between 10:30 pm and 5:30 am, following a common approach in the literature \citep{athey2021estimating, cook2024urban, moro2021mobility}. We only keep a phone in a given month if it has five or more pings per month at the home location. For the entire year, there are 746k unique cell phone identifiers in Bogota for which we find a home location, and these have an average of 451 pings and 82 trips per month. To detect trips, we use the \cite{hariharan2004toyama} algorithm which is a common methodology for trip detection \citep[see e.g.][]{moro2021mobility}. We define a trip as a series of location signals within a radius of 100 metres, a minimum duration of 5 minutes, and a maximum duration of 24 hours. We use these trips to measure the mobility of the treatment and the control group, for example, by calculating the number of trips outside the home an individual makes per month. We also calculate further metrics, such as the number of trips to specific areas of Bogota.  We also match trips to specific POIs (if there is no POI within a radius of 50 meters of a trip destination, then the trip is not matched to a POI). 

\textbf{Metrics for experienced segregation.} To understand how the opening of the cable car affects the exposure of the treatment group to individuals from different socioeconomic groups, we use two metrics to measure the exposure of low-income individuals to higher-income individuals when they visit POIs. First, the Shannon entropy index measures the evenness of visits to a POI across stratum categories (by all Bogota residents with a home location in our dataset). Second, we also compute the share of visits to a POI made by high-income individuals, who we define as stratum categories 4 (upper-middle-income), 5 (high-income), or 6 (very high-income).  

For both measures, in a first step, we calculate the Shannon entropy/share of high-income visits for all POIs. For example, we calculate the Shannon entropy of the trips made to a specific park (for all Bogota residents). In a second step, we calculate the experienced segregation at the individual-level for individuals living in the treatment or the control group over time. In each month, we calculate the average Shannon entropy/share of high-income visits of the POIs visited by individuals in the treatment and the control group. When doing so, we hold constant the levels of experienced segregation of the POIs. Therefore, we assume that changes in the visits of an individual do not alter the overall patterns of experienced segregation of a given POI. We make this assumption, which is common in the literature \citep{athey2021estimating, moro2021mobility}, as individuals in the treatment and control group only make up a small fraction (only 1\%) of all individuals observed. 

\begin{footnotesize}
\bibliographystyle{agsm}
\bibliography{Refs.bib}
\end{footnotesize}

\section*{Supplementary Information}

Coming soon.

\end{document}